\documentclass[prd,aps,floatfix,onecolumn,showpacs,nofootinbib]{revtex4}

\usepackage[dvipdfm]{graphicx}
\usepackage{bm}
\usepackage{amsmath,amssymb}
\usepackage{natbib}

\input{colordvi.tex}

\newcommand{\simgt}{\lower.5ex\hbox{$\; \buildrel > \over \sim \;$}}
\newcommand{\simlt}{\lower.5ex\hbox{$\; \buildrel < \over \sim \;$}}

\everymath{\displaystyle}
\def\pd{{\rm d}}

\def\reff@jnl#1{{\rm#1\/}}
\def\aj{\reff@jnl{AJ}}                  
\def\araa{\reff@jnl{ARA\&A}}            
\def\apj{\reff@jnl{ApJ}}                
\def\apjl{\reff@jnl{ApJ}}               
\def\apjs{\reff@jnl{ApJS}}              
\def\ao{\reff@jnl{Appl.Optics}}         
\def\apss{\reff@jnl{Ap\&SS}}            
\def\aap{\reff@jnl{A\&A}}               
\def\aapr{\reff@jnl{A\&A~Rev.}}         
\def\aaps{\reff@jnl{A\&AS}}            
\def\azh{\reff@jnl{AZh}}                
\def\baas{\reff@jnl{BAAS}}              
\def\jrasc{\reff@jnl{JRASC}}            
\def\memras{\reff@jnl{MmRAS}}           
\def\mnras{\reff@jnl{MNRAS}}            
\def\newastro{\reff@jnl{New Astron.}}   
\def\pra{\reff@jnl{Phys.Rev.A}}         
\def\prb{\reff@jnl{Phys.Rev.B}}         
\def\prc{\reff@jnl{Phys.Rev.C}}         
\def\prd{\reff@jnl{Phys.Rev.D}}         
\def\prl{\reff@jnl{Phys.Rev.Lett}}      
\def\pasp{\reff@jnl{PASP}}              
\def\pasj{\reff@jnl{PASJ}}              
\def\qjras{\reff@jnl{QJRAS}}            
\def\skytel{\reff@jnl{S\&T}}            
\def\solphys{\reff@jnl{Solar~Phys.}}    
\def\sovast{\reff@jnl{Soviet~Ast.}}     
\def\ssr{\reff@jnl{Space~Sci.Rev.}}     
\def\zap{\reff@jnl{ZAp}}                
\def\nat{\reff@jnl{Nature}}             
\def\physrep{\reff@jnl{Phys.~Rep.}}     
\def\prog{\reff@jnl{PThPS}}        

\begin{document}

\title{Copula Cosmology: Constructing a Likelihood Function }

\author{Masanori Sato$^{1,2}$\footnote{masanori@a.phys.nagoya-u.ac.jp}, Kiyotomo Ichiki$^{1}$, Tsutomu T. Takeuchi$^{3}$}
\affiliation{%
$^{1}$ Department of Physics, Nagoya University, Nagoya 464--8602, Japan
}%
\affiliation{%
$^{2}$ Lawrence Berkeley National Laboratory, 1 Cyclotron Road,
Berkeley, California 94720, USA
}%
\affiliation{%
$^{3}$ Institute for Advanced Research, Nagoya University, Nagoya
464--8601, Japan}%

\date{\today}

\begin{abstract}
{}To estimate cosmological parameters from a given dataset, we need to
construct a likelihood function, which sometimes has a complicated functional form.
We introduce the copula, a mathematical tool to construct an arbitrary
multivariate distribution function from one-dimensional marginal
distribution functions with any given dependence structure.  
It is shown that a likelihood function constructed by the so-called Gaussian 
copula can reproduce very well the $n$-dimensional probability distribution of the
cosmic shear power spectrum obtained from a large number of ray-tracing
simulations. 
This suggests that the Copula likelihood will be a powerful tool for future weak 
lensing analyses, instead of the conventional
multivariate Gaussian likelihood.
\end{abstract}

\maketitle

\section{Introduction}
Weak gravitational lensing by intervening large scale cosmic structures provides
an excellent tool to probe the nature of dark matter and dark energy.
The so-called ``cosmic shear'' signal has been successfully measured in
various groups since 2000
\citep[e.g.][]{2000MNRAS.318..625B,2000astro.ph..3338K,2000A&A...358...30V,2000Natur.405..143W,2003ApJ...597...98H,2006ApJ...644...71J,2006A&A...452...51S,2008A&A...479....9F,2010A&A...516A..63S,2011MNRAS.410..143S}.
If systematic errors are well under control, the weak lensing has the
highest potential to constrain the physical parameters in the equation of state governing 
the dark energy among the cosmological
observations, such as type Ia supernovae, baryon acoustic oscillations,
and number count of galaxy clusters
\citep{2006astro.ph..9591A,2009arXiv0901.0721A,2009PhRvD..80b3003J}.

A number of wide-field weak lensing surveys have been planned for this
purpose, such as the Subaru Hyper Suprime-Cam
Survey~\citep{2006SPIE.6269E...9M}, the Panoramic Survey Telescope \&
Rapid Response
System~(Pan-STARRS\footnote{http://pan-starrs.ifa.hawaii.edu/public/}),
the Dark Energy
Survey~(DES\footnote{http://www.darkenergysurvey.org/}), the Large
Synoptic Survey Telescope~(LSST\footnote{http://www.lsst.org/}), the
Joint Dark Energy Mission~(JDEM\footnote{http://jdem.gsfc.nasa.gov/})
and Euclid~\citep{2010arXiv1001.0061R}.  
It is expected that such wide-field weak lensing surveys will reduce {\it statistical}
errors significantly compared to the past and ongoing surveys, because the number of
observed galaxies increases proportional to the survey area.  
However, to make a maximal use of the full potential of planned weak lensing surveys for
estimating cosmological parameters, it is of great importance to employ
adequate statistical measures and methods for weak lensing.  
Particularly, one needs to take into account properly the correlations of the observables
between different angular scales and redshifts, i.e. the covariances.
Furthermore, we also need to use an appropriate likelihood function with given marginal distributions.  
If we do not use proper statistical measures and methods, or if we adopt an inaccurate
covariance and/or a likelihood function, obtained results may be
systematically biased \citep{2009A&A...504..689H,2009PhRvD..79b3520I}.

For the cosmological parameter estimation, almost all previous authors used
the $\chi^2$ method in weak lensing analyses
\citep[e.g.][]{2003ApJ...597...98H,2006ApJ...644...71J,2006A&A...452...51S,2008A&A...479....9F,2010A&A...516A..63S,2011MNRAS.410..143S}.
However, it is found that the probability
distribution function (PDF) of the weak lensing power spectrum is well
approximated by the $\chi^2$ distribution though it has a larger
positive tail than expected from the $\chi^2$ distribution \cite{2009ApJ...701..945S}.  
The $\chi^2$ distribution deviates from the Gaussian distribution on large
scales because the number of modes corresponding to the degree of
freedom is very small.  
Meanwhile, the $\chi^2$ distribution approaches the Gaussian distribution at high 
$\ell$ due to the central limit theorem.
We have to include this information accurately when we place constraints
on the cosmological parameters.

If all of the marginal distributions are Gaussian distributed, it is
straightforward to reconstruct the multivariate PDF, which is the so-called
multivariate Gaussian PDF.  
However, it is not a trivial task to reconstruct the original multivariate PDF from 
general marginal distributions.  
There has been an infinite number of the degree of
freedom in choosing the original PDF unless the dependence structure is specified.  
The copula provides us with a straightforward solution to this problem.
The copula has been used in the field of mathematical finance, but not been widely used 
in the field of astronomy and cosmology, except only a few applications
 \citep{2009MNRAS.400..219B,2009AJ....137..329J,2009MNRAS.393.1370K,2010ApJ...708L...9S,2010MNRAS.406.1830T}.
Hence, a copula may have a potential to open new fields of astronomy and cosmology.

In this paper, we construct a more plausible likelihood function using
the Gaussian copula (hereafter ``Copula likelihood'') than the multivariate
Gaussian likelihood for the cosmic shear power spectrum.  
We show that the Copula likelihood well reproduces the $n$-dimensional probability
distribution of the cosmic shear power spectrum estimated from 1000
realizations which is obtained from ray-tracing simulations performed by
\cite{2009ApJ...701..945S}.  
Cosmological parameters employed for our ray-tracing simulations are consistent 
with the WMAP 3-year results \citep{2007ApJS..170..377S}.  
The detail descriptions of our ray-tracing simulations are summarized in
\cite{2009ApJ...701..945S} (see also, \cite{2010arXiv1009.2558S}).  
In a companion paper \citep{2010PhRvL.105y1301S}, we estimate the cosmological parameters
using both the copula likelihood constructed by this paper and the
Gaussian likelihood in order to evaluate how the difference between two
likelihoods affects the parameter estimation.

\section{Formulation}
The likelihood function is a central tool for any kind of parameter estimation.
It is defined as a function of parameters in a given statistical model.
The likelihood function $\mathcal{L}$ is
related to the joint probability density function (JPDF) denoted by
$f(\hat{x}_{1},\hat{x}_{2},\dots,\hat{x}_{n}\,|\,\theta)$ as 
\begin{equation}
 \mathcal{L}(\theta\,|\,\hat{x}_1,\hat{x}_2,\dots,\hat{x}_n)=f(\hat{x}_1,\hat{x}_2,\dots,\hat{x}_n\,|\,\theta),
\end{equation}
where $\hat{x}_i$ ($i=1,2,\dots,n$) are independent and
identically-distributed observed variables and $\theta$ denotes model parameters.
We will suppress the argument $\theta$ hereafter.
Recall that the $n$-point cumulative distribution function (CDF),
denoted by $F$, is
defined as 
\begin{equation}
 F(\hat{x}_1,\hat{x}_2,\dots,\hat{x}_n)=\int_{-\infty}^{\hat{x}_1}\int_{-\infty}^{\hat{x}_2}\dots\int_{-\infty}^{\hat{x}_n}f(x_1,x_2,\dots,x_n)\pd
  x_1\pd x_2\dots \pd x_n,
\end{equation}
and there is a following relation between JPDF and CDF:
\begin{equation}
 f(\hat{x}_1,\hat{x}_2,\dots,\hat{x}_n)=\frac{\partial^n
 F(\hat{x}_1,\hat{x}_2,\dots,\hat{x}_n)}{\partial\hat{x}_1\partial\hat{x}_2\dots\partial\hat{x}_n}.
\label{cdfpdf_rel}
\end{equation}

From the Sklar's theorem~\citep{sklar1959fonctions}, we can obtain the following relation:
\begin{equation}
 {\rm
  Prob}(x_{1}\le\hat{x}_{1},x_{2}\le\hat{x}_{2},\dots,x_{n}\le\hat{x}_{n})\equiv
  F(\hat{x}_{1},\hat{x}_{2},\dots,\hat{x}_{n})=C(F_1(\hat{x}_{1}),F_2(\hat{x}_{2}),\dots,
  F_n(\hat{x}_{n})),
\label{sklars_theorem}
\end{equation}
where $C$ denotes the function called copula and $F_{i}$ denotes
 one-point CDF defined by
\begin{equation}
 F_i(\hat{x}_i)=\int_{-\infty}^{\hat{x}_i}f_i(x)\pd x\equiv u_i.
\end{equation}
Therefore, the copula indicates how the one-point CDFs are jointed together to 
give the $n$-point CDF.
A comprehensive proof of Sklar's theorem and rigorous definition of a
copula are found in \cite{nelsen2006introduction,2010MNRAS.406.1830T}.
{}From the above relation, we can easily derive that $\hat{x}_i=F^{-1}_i(u_i)$,
and then derive the following relation,
\begin{equation}
 C(u_1,u_2,\dots,u_n)=F(F_1^{-1}(u_1),F_2^{-1}(u_2),\dots,F_n^{-1}(u_n))~,
\label{copula_1}
\end{equation}
{}from Eq.~(\ref{sklars_theorem}).
Differentiating Eq.~(\ref{copula_1}) with Eq.~(\ref{cdfpdf_rel}) 
gives the density of copula $c$ as
\begin{equation}
 c(u_1,u_2,\dots,u_n)=\frac{\partial^n C(u_1,u_2,\dots,u_n)}{\partial
  u_1\partial u_2\dots\partial
  u_n}=\frac{f(\hat{x}_1,\hat{x}_2,\dots,\hat{x}_n)}{\displaystyle\prod_{i=1}^{n}f_i(\hat{x}_i)},
\label{dens_copu}
\end{equation}
where each $f_i$ is the marginal density function of the marginal CDF
$F_i$.
The JPDF can then be expressed as
\begin{equation}
 f(\hat{x}_1,\hat{x}_2,\dots,\hat{x}_n)=c(u_1,u_2,\dots,u_n)\prod_{i=1}^n
  f_i(\hat{x}_i)\label{formu:f}.
\end{equation}
If the variables $x_i$ are independent of each other, $c=1$. 
In general, however, they are often correlated and $c$ shows their
correlations as a function of one-point CDF of each stochastic variable.

\section{Gaussian Copula}
In this section, we derive the Copula likelihood using a Gaussian copula
which is more plausible than the multivariate Gaussian likelihood
for the cosmic shear power spectrum.  
The multivariate Gaussian copula
is the copula of the $n$-dimensional random vector that is normally
distributed.  
This copula is expressed by
\begin{equation}
 C(u_1,u_2,\dots,u_n)\equiv\Phi\left(\Phi_1^{-1}(u_1),\Phi_1^{-1}(u_2),\dots,\Phi_1^{-1}(u_n)\right).
\end{equation}
Here one-point Gaussian CDF $\Phi_1$ is 
\begin{equation}
 \Phi_1(\hat{x}_i)=\int_{-\infty}^{\hat{x}_i}\frac{1}{\sqrt{2\pi}\sigma_i}\exp\left(-\frac{(x-\mu_i)^2}{2\sigma_i^2}\right)\pd x\equiv
  u_i,
\end{equation}
and $\Phi$ is $n$-point Gaussian CDF defined by
\begin{equation}
 \Phi(\hat{x}_1,\hat{x}_2,\dots,\hat{x}_n)=\int_{-\infty}^{\hat{x}_1}\int_{-\infty}^{\hat{x}_2}\dots\int_{-\infty}^{\hat{x}_n}\frac{1}{\sqrt{(2\pi)^n
  {\rm
  det(Cov)}}}\exp\left(-\frac{1}{2}(\boldsymbol{x}-\boldsymbol{\mu})^{\rm T}{\rm
 Cov}^{-1}(\boldsymbol{x}-\boldsymbol{\mu})\right)\pd x_1\pd x_2\dots
  \pd x_n,
\end{equation}
where we consider the $n$-point Gaussian CDF with mean $\boldsymbol{\mu}$ and
$n\times{n}$ covariance matrix.
${\rm Cov}^{-1}$ shows the inverse covariance matrix.
We define $\boldsymbol{\mu}\equiv (\mu_1,\mu_2,\dots,\mu_n)$ and
$\boldsymbol{x}\equiv (x_1,x_2,\dots,x_n)$ and superscript 'T'
stands for the transpose of vector.

\subsection{Gaussian PDF}
Let us consider a two-variables case for simplicity.
First consider the case that one-point PDF is Gaussian
with $\sigma_i$ standard deviation and $\mu_i$ mean.
In this case, the Gaussian copula is 
\begin{align}
&C(\hat{x}_1,\hat{x}_2)=\Phi\left(\Phi_1^{-1}(\Phi_1(\hat{x}_1)),\Phi_1^{-1}(\Phi_1(\hat{x}_2))\right),\label{ex:copula}\\
&\Phi(\hat{x}_1,\hat{x}_2)=\int_{-\infty}^{\hat{x}_1}\int_{-\infty}^{\hat{x}_2}
\frac{1}{\sqrt{(2\pi)^2 {\rm
 det(Cov)}}}\exp\left(-\frac{1}{2}(\boldsymbol{x}-\boldsymbol{\mu})^{\rm
 T}{\rm
 Cov}^{-1}(\boldsymbol{x}-\boldsymbol{\mu})\right)\pd x_1\pd
 x_2\equiv\int_{-\infty}^{\hat{x}_1}\int_{-\infty}^{\hat{x}_2}\phi(x_1,x_2)\pd x_1\pd x_2.\label{ex:phi}
\end{align}
{}From Eq.~(\ref{dens_copu}), we derive the density of copula as
\begin{align}
 c(u_1,u_2)=\frac{\partial\Phi(\hat{x}_1,\hat{x}_2)}{\partial
 u_1\partial u_2}=\frac{\partial^2\Phi}{\partial \hat{x}_1\partial \hat{x}_2}\frac{\partial
 \hat{x}_1}{\partial u_1}\frac{\partial \hat{x}_2}{\partial u_2}=\phi(\hat{x}_1,\hat{x}_2)\frac{\partial
 \hat{x}_1}{\partial u_1}\frac{\partial \hat{x}_2}{\partial u_2}.
\end{align}
By using 
\begin{align}
\frac{\partial u_i}{\partial
\hat{x}_i}=\frac{1}{\sqrt{2\pi}\sigma_i}\exp\left(-\frac{(\hat{x}_i-\mu_i)^2}{2\sigma_i^2}\right),
\end{align}
we finally obtain the density of copula as
\begin{align}
  c(u_1,u_2)&=\frac{\frac{1}{\sqrt{(2\pi)^2 {\rm
 det(Cov)}}}\exp\left(-\frac{1}{2}(\boldsymbol{\hat{x}}-\boldsymbol{\mu})^{\rm
 T}{\rm
 Cov}^{-1}(\boldsymbol{\hat{x}}-\boldsymbol{\mu})\right)}{\frac{1}{\sqrt{2\pi}\sigma_1}\exp\left(-\frac{(\hat{x}_1-\mu_1)^2}{2\sigma_1^2}\right)\frac{1}{\sqrt{2\pi}\sigma_2}\exp\left(-\frac{(\hat{x}_2-\mu_2)^2}{2\sigma_2^2}\right)}\nonumber\\
&=\frac{\sigma_1\sigma_2}{\sqrt{\rm
 det(Cov)}}\exp\left[-\frac{1}{2}\left[(\boldsymbol{\hat{x}}-\boldsymbol{\mu})^{\rm
 T}({\rm
 Cov}^{-1}-(\boldsymbol{I\sigma^2})^{-1})(\boldsymbol{\hat{x}}-\boldsymbol{\mu})\right]\right],
\end{align}
where $\boldsymbol{I}$ stands for the identity matrix.
Therefore, we can obtain the JPDF as
\begin{align}
  f(\hat{x}_1,\hat{x}_2)&=\frac{\sigma_1\sigma_2}{\sqrt{\rm
 det(Cov)}}\exp\left[-\frac{1}{2}\left[(\boldsymbol{\hat{x}}-\boldsymbol{\mu})^{\rm
 T}({\rm
 Cov}^{-1}-(\boldsymbol{I\sigma^2})^{-1})(\boldsymbol{\hat{x}}-\boldsymbol{\mu})\right]\right]\prod_{i=1}^{2}\frac{1}{\sqrt{2\pi}\sigma_i}\exp\left(-\frac{(\hat{x}_i-\mu_i)^2}{2\sigma_i^2}\right)\nonumber\\
&=\frac{1}{\sqrt{(2\pi)^2 {\rm det(Cov)}}}\exp\left(-\frac{1}{2}(\boldsymbol{\hat{x}}-\boldsymbol{\mu})^{\rm
 T}{\rm Cov}^{-1}(\boldsymbol{\hat{x}}-\boldsymbol{\mu})\right)
\end{align}
{}from Eq.~(\ref{formu:f}).
Therefore, the JPDF using the Gaussian copula with a Gaussian one-point
PDF results in the multivariate Gaussian distribution, as expected.

\subsection{Beyond Gaussian PDF}
Now, let us consider the case in which a one-point PDF is not a Gaussian
distribution but a general probability distribution $f_i$.
In this case, the Gaussian copula is 
\begin{equation}
 C(\hat{x}_1,\hat{x}_2)=\Phi\left(\Phi_1^{-1}(F_1(\hat{x}_1)),\Phi_1^{-1}(F_2(\hat{x}_2))\right),\label{be:copula}
\end{equation}
where one-point CDF $F_i$ is 
\begin{equation}
 F_i(\hat{x}_i)=\int_{-\infty}^{\hat{x}_i}f_i(x)\pd x\equiv
  u_i.\label{be:u_i}
\end{equation}
Defining $q_i$ as
$q_i\equiv\Phi_1^{-1}(F_i(\hat{x}_i))=\Phi_1^{-1}(u_i)$, we obtain the
copula density from
Eq.~(\ref{be:copula}) as 
\begin{equation}
 c(u_1,u_2)=\frac{\partial^2\Phi(q_1,q_2)}{\partial u_1\partial
  u_2}=\frac{\partial^2\Phi}{\partial q_1\partial q_2}\frac{\partial
  q_1}{\partial u_1}\frac{\partial q_2}{\partial u_2}\label{be:scopula},
\end{equation}
where
\begin{align}
 \frac{\partial
  q_i}{\partial u_i}=\frac{\partial\Phi_1^{-1}(u_i)}{\partial
  u_i}=\left(\frac{\partial\Phi_1(q_i)}{\partial q_i}\right)^{-1}=\left(\frac{1}{\sqrt{2\pi}\sigma_i}\exp\left(-\frac{(q_i-\mu_i)^2}{2\sigma_i^2}\right)\right)^{-1}.\label{be:diff}
\end{align}
We use a formula of the differential of the inverse function at the second
equality of the above equation.
By using Eq.~(\ref{ex:phi}) and Eq.~(\ref{be:diff}), we rewrite Eq.~(\ref{be:scopula}) as
\begin{align}
 c(u_1,u_2)=\phi(q_1,q_2)\left(\frac{1}{\sqrt{2\pi}\sigma_1}\exp\left(-\frac{(q_1-\mu_1)^2}{2\sigma_1^2}\right)\right)^{-1}\left(\frac{1}{\sqrt{2\pi}\sigma_2}\exp\left(-\frac{(q_2-\mu_2)^2}{2\sigma_2^2}\right)\right)^{-1}\label{be:scopula_v1}.
\end{align}
{}From Eq.~(\ref{formu:f}), we can easily calculate the JPDF as
\begin{align}
 f(\hat{x}_1,\hat{x}_2)&=\frac{1}{\sqrt{(2\pi)^2 {\rm
 det(Cov)}}}\exp\left(-\frac{1}{2}(\boldsymbol{q}-\boldsymbol{\mu})^{\rm
 T}{\rm
 Cov}^{-1}(\boldsymbol{q}-\boldsymbol{\mu})\right)\left(\frac{1}{\sqrt{2\pi}\sigma_1}\exp\left(-\frac{(q_1-\mu_1)^2}{2\sigma_1^2}\right)\right)^{-1}\nonumber\\
&\times\left(\frac{1}{\sqrt{2\pi}\sigma_2}\exp\left(-\frac{(q_2-\mu_2)^2}{2\sigma_2^2}\right)\right)^{-1}f_1(\hat{x}_1)f_2(\hat{x}_2).
\end{align}
We extend the above equation to the $n$-dimensional case and then
finally obtain the $n$-dimensional JPDF as
\begin{align}
 f(\hat{x}_1,\hat{x}_2,\dots,\hat{x}_n)&=\frac{1}{\sqrt{(2\pi)^n {\rm
 det(Cov)}}}\exp\left(-\frac{1}{2}(\boldsymbol{q}-\boldsymbol{\mu})^{\rm
 T}{\rm
 Cov}^{-1}(\boldsymbol{q}-\boldsymbol{\mu})\right)\prod_{i=1}^{n}\left(\frac{1}{\sqrt{2\pi}\sigma_i}\exp\left(-\frac{(q_i-\mu_i)^2}{2\sigma_i^2}\right)\right)^{-1}f_i(\hat{x}_i).
\label{be:copula_f}
\end{align}
If $f_i$ is Gaussian, $q_i$ becomes $\hat{x}_i$.
In this case, we can see that $f(\hat{x}_1,\hat{x}_2,\dots,\hat{x}_n)$
(Eq.~\ref{be:copula_f}) reduces to a multivariate Gaussian distribution.
Therefore, when $f_i$ are not Gaussian PDFs, Eq.~(\ref{be:copula_f}) carries 
the whole information on the correction to the Gaussian distribution.

Both in practice and from theoretical point of view
\citep{2000Ap&SS.271..213T}, 
it is more appropriate to work with the logarithm of
the likelihood function, $\ln\mathcal{L}$, called log-likelihood.
For example, the log-likelihood is used for the likelihood ratio test.
The likelihood ratio test statistic is defined as twice the difference in
these log-likelihoods with a minus sign.
This quantity is also fundamental for the information statistics and 
information criterion theory (see, e.g. \citep{2000Ap&SS.271..213T} and
references therein).
{}From Eq.~(\ref{be:copula_f}), this test statistic is derived as
\begin{align}
 -2\ln\mathcal{L}_{c}(\hat{x}_1,\hat{x}_2,\dots,\hat{x}_n)=
\sum_{i=1}^{n}\sum_{j=1}^n(q_i-\mu_i){\rm
 Cov}^{-1}(q_j-\mu_j)-\sum_{i=1}^n\frac{(q_i-\mu_i)^2}{\sigma_i^2}-2\sum_{i=1}^n\ln f_i(\hat{x}_i),
\label{gau_copula_like}
\end{align}
for a general probability distribution. For a Gaussian case it reduces
to the well-known form
\begin{equation}
 -2\ln\mathcal{L}_{g}(\hat{x}_1,\hat{x}_2,\dots,\hat{x}_n)=
\sum_{i=1}^{n}\sum_{j=1}^n(\hat{x}_i-\mu_i){\rm
 Cov}^{-1}(\hat{x}_j-\mu_j).
\label{gau_like}
\end{equation}
Here we abbreviate the irrelevant constant term in the above two equations.

Finally, we give the relation between $\hat{x}_i$ and $q_i$ because one has
to calculate $q_i$ given $\hat{x}_i$.
Since $\hat{x}_i$ is related to $u_i$ through Eq.~(\ref{be:u_i}), we have
 only to derive the relation between $u_i$ and $q_i$.
From $q_i=\Phi_1^{-1}(u_i)$, we get 
\begin{equation}
 u_i=\Phi_1(q_i)=\int_{-\infty}^{q_i}\frac{1}{\sqrt{2\pi}\sigma_i}\exp\left(-\frac{(x-\mu_i)^2}{2\sigma_i^2}\right)\pd x.
\end{equation}
We change the variable $x$ to $y=(x-\mu_i)/\sigma_i$ and then obtain
\begin{equation}
 u_i=\frac{1}{\sqrt{2\pi}}\int_{-\infty}^{\frac{q_i-\mu_i}{\sigma_i}}\exp\left(-\frac{y^2}{2}\right)\pd{y}.
\end{equation}
At a first glance, we can recognize this equation as 
$u_i=\Psi_1\left(\frac{q_i-\mu_i}{\sigma_i}\right)$, where $\Psi_1$ is the
cumulative standard normal distribution.
We can obtain the relation between $u_i$ and $q_i$ as  
\begin{equation}
 \frac{q_i-\mu_i}{\sigma_i}=\Psi_1^{-1}(u_i)\Leftrightarrow q_i=\sigma_i\Psi_1^{-1}(u_i)+\mu_i.
\end{equation}

\section{Result and Discussion}
In this section, we investigate whether the Copula likelihood (Eq.~\ref{gau_copula_like})
reproduces the $n$-dimensional JPDF obtained from ray-tracing
simulations performed by \cite{2009ApJ...701..945S}.  We then compare
the shape of the Copula likelihood with that of the
Gaussian likelihood and how the Copula likelihood is {\em
statistically} better than the Gaussian likelihood. 
In our application, the observed variables
$\hat{x}_i$ are the binned convergence power spectra
$\hat{P}_{\kappa}(\ell_i)$, which are estimated from each realization.  We
estimate it for an assumed bin width $\Delta\ln{\ell}$.  Throughout this
paper, we employ the bin width $\Delta\ln \ell=0.3$ and assume that the
single source redshift distribution, i.e. all lensed galaxies lie at
$z_s=1.0$ and we do not consider intrinsic ellipticity dispersion 
$\sigma_{\epsilon}$.
In our ray-tracing simulations, the survey area is set as $\Omega_{\rm
s}=25$ deg$^2$.
Therefore the fundamental mode of our ray-tracing simulations is at the 
multipole $\ell_f=72$.  We take 13 bins for our likelihood analysis.
Therefore, we cover the multipole range from $\ell_f=72$ to $\ell_{\rm
max}=2635$.  We call the convergence power spectrum estimated
at bin $i$ as ``bin $i$ power''.
 
\begin{figure}[!t]
\begin{minipage}{.48\textwidth}
\begin{center}
\includegraphics[width=0.95\textwidth]{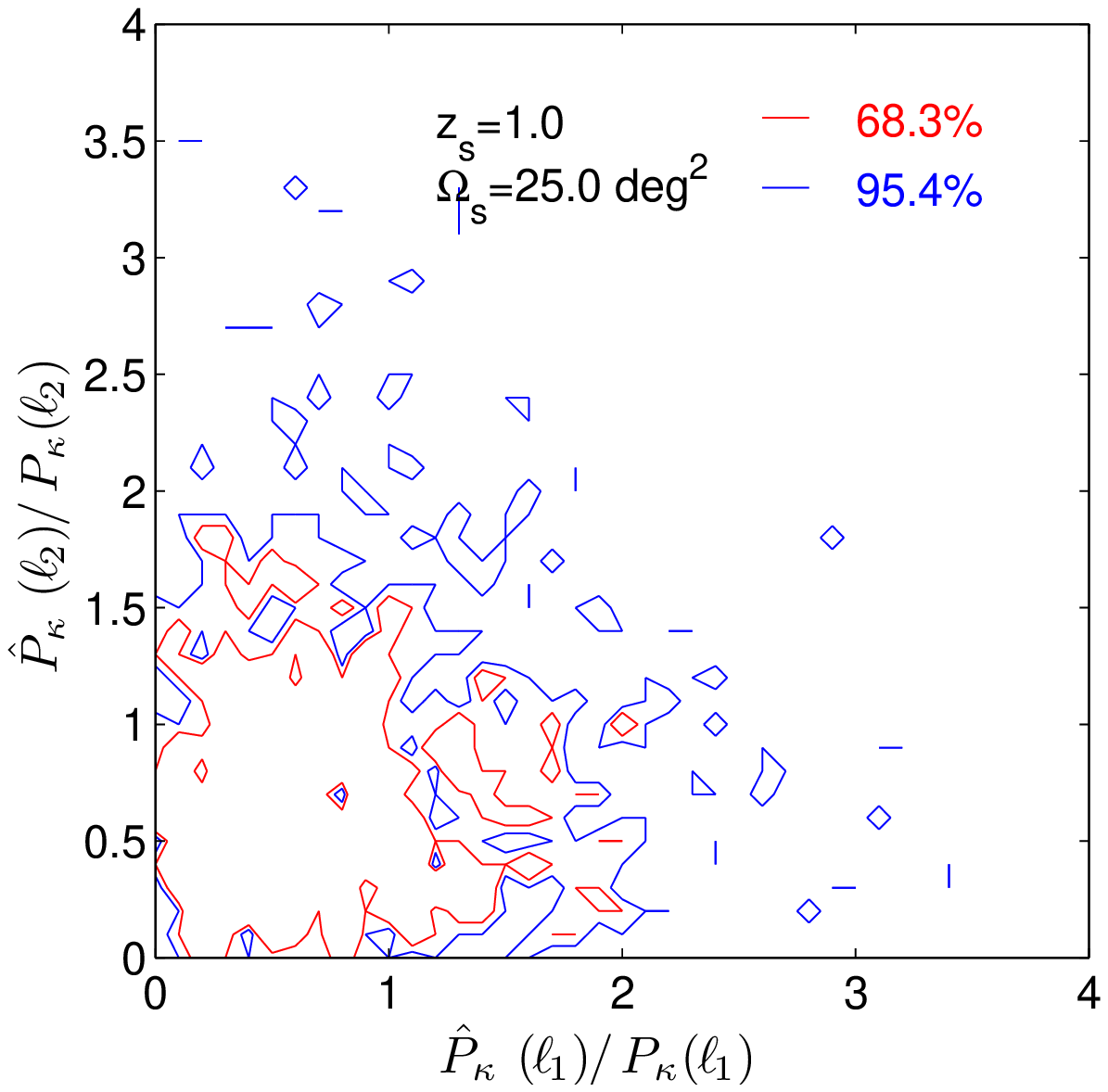}
\end{center}
\end{minipage}
\begin{minipage}{.48\textwidth}
\begin{center}
\includegraphics[width=0.95\textwidth]{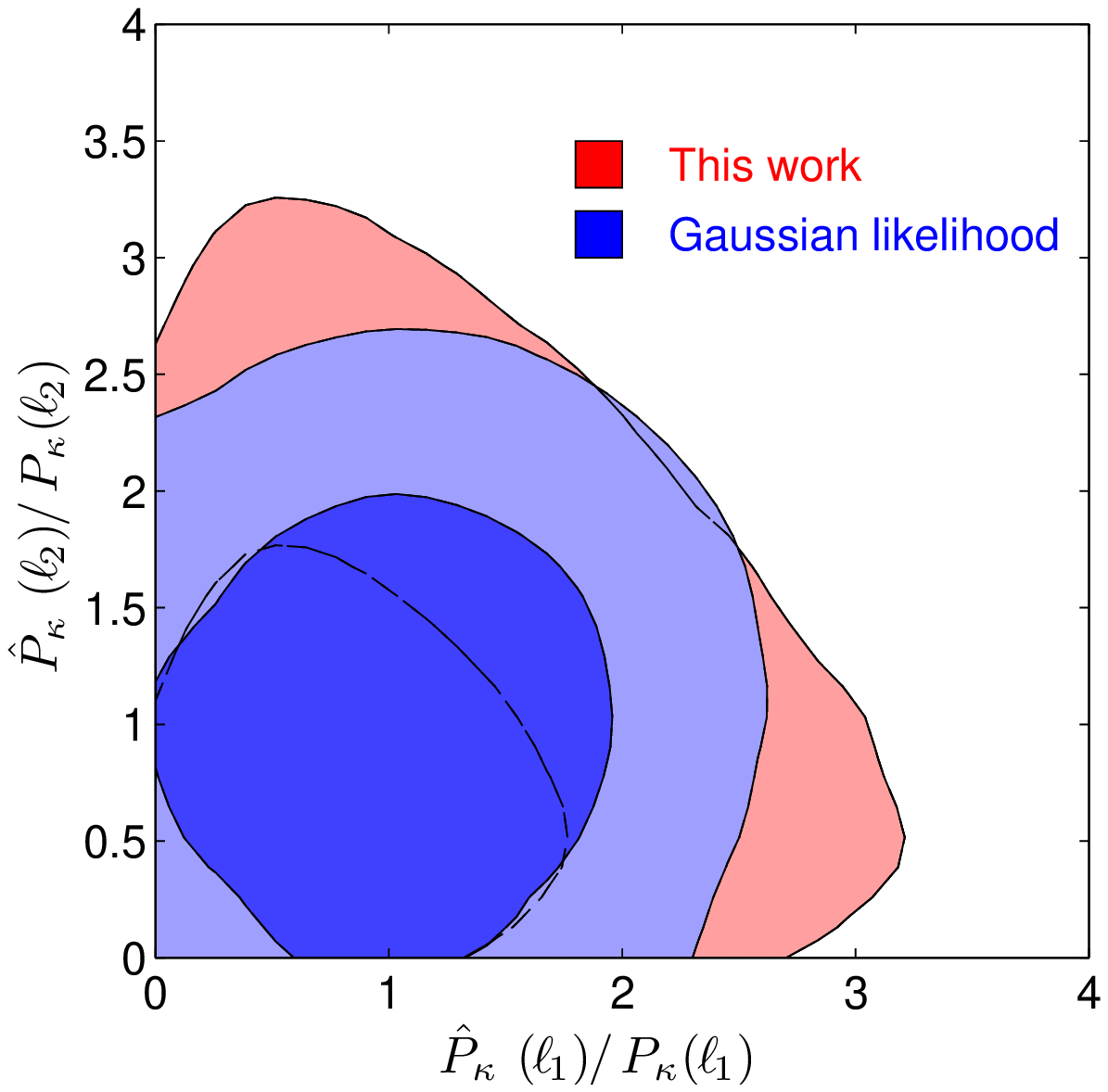}
\end{center}
\end{minipage}
\vskip-\lastskip
\caption{{\it Left panel}: Two-dimensional JPDF between convergence power
 spectrum estimated at bin 1 and bin 2 among 1000 realizations
 obtained from ray-tracing simulations.
 The bin 1 and bin 2 correspond to multipole $\ell=72$ and $\ell=97$,
 respectively.
 The one-point PDF is normalized so that the mean convergence power spectrum
 estimated at each bin gives unity. 
 The red solid and blue solid contours show 1$\sigma$ and
 2$\sigma$ CL, respectively.
 {\it Right panel}: Two-dimensional marginalized constraints on
 convergence power spectrum estimated at bin 1 and bin 2.
 The red and blue contours show the marginalized constraints (1$\sigma$ and
 2$\sigma$ CL) obtained by Eq.~(\ref{gau_copula_like}) based on
 the Gaussian copula model and Gaussian likelihood of
 Eq.~(\ref{gau_like}), respectively.}
\label{fig:2dim0102}
\end{figure}

The left panel of Fig.~\ref{fig:2dim0102} shows the two-dimensional JPDF between
bin 1 power and bin 2 power from 1000 realizations of ray-tracing
simulations.  The one-point PDF is normalized so that the mean power
spectrum is equal to unity.  The red solid and blue solid contours show
the 1$\sigma$ and 2$\sigma$ confidence level (CL) regions, respectively.
In the right panel of Fig.~\ref{fig:2dim0102}, the blue and red contours
show two-dimensional marginalized 1$\sigma$ and 2$\sigma$ confidence
regions on the bin 1 power and bin 2 power plane, which
are derived from the Gaussian likelihood (Eq.~\ref{gau_like}) and
Copula likelihood (Eq.~\ref{gau_copula_like}), respectively. One can
clearly see that the Copula likelihood model reproduces the simulation
data (left panel of Fig.~\ref{fig:2dim0102}) much better than the
Gaussian likelihood model, as discussed below.

In our Copula likelihood model, we have chosen a $\chi^2$ distribution for
a general one-point probability distribution $f_i$.
This is a nice choice because the one-point PDF of convergence power spectrum is fairly
well described by a $\chi^2$ distribution where the mean and variance are
$P_{\kappa}(\ell_i)=\langle{\hat{P}_{\kappa}(\ell_i)}\rangle$ and
$\sigma^2(\ell_i)=\langle\hat{P}_{\kappa}(\ell_i)^2\rangle -
P_{\kappa}(\ell_i)^2$, respectively \citep[see][]{2009ApJ...701..945S}.
From \cite{2009ApJ...700..479T}, this $\chi^2$ distribution is shown as
\begin{equation}
 f_{\chi^2}(\hat{P}_{\kappa}(\ell_i))=\frac{\hat{P}_{\kappa}(\ell_i)^{\Upsilon
  -1}}{\Gamma(\Upsilon)}\left(\Upsilon\frac{e^{-\hat{P}_{\kappa}(\ell_i)/P_{\kappa}(\ell_i)}}{P_{\kappa}(\ell_i)}\right)^{\Upsilon}, \label{Eq:chi2dist}
\end{equation}
for $\hat{P}_{\kappa}(\ell_i)>0$ and $f_{\chi^2}=0$ for
$\hat{P}_{\kappa}(\ell_i)\le 0$.
Here, $\Gamma(x)$ is the gamma function and we define
$\Upsilon\equiv{P_{\kappa}(\ell_i)^2/\sigma^2(\ell_i)}$ which
corresponds to the number of independent modes. The covariance matrix is
also estimated from the simulation.

In order to obtain these likelihood contours, we employed the Markov
Chain Monte Carlo method \citep{2002PhRvD..66j3511L}.
Assuming flat priors for bin $i$ powers, we explored bin $i$ power estimations
in the multidimensional space (i.e. 13 dimensional space in this case).
Eight parallel chains are computed and the convergence test is made
based on the Gelman and Rubin statistics called ``$R-1$'' statistics
\citep{gelman1992inference}.
Each of our chains typically has 400,000 points and $R-1<0.05$ in both models.

We can see that the results between our Copula likelihood function (red
contours) and the Gaussian likelihood (blue contours) are very
different.  The difference mainly comes from the one-point PDF, which is
taken as a $\chi^2$ distribution or a Gaussian distribution.  The
$\chi^2$ distribution denoted by Eq.~(\ref{Eq:chi2dist}) deviates from
the Gaussian distribution on large scales, such as a case considered in
Fig.~\ref{fig:2dim0102}, because the number of modes corresponding to
the degrees of freedom is very small.  
We can also see that our likelihood
function well reproduces the results from ray-tracing simulations than
the Gaussian likelihood.  In particular, the results of our Copula
likelihood function capture the feature that the value which takes the
maximum probability deviates from the mean value.
The values at which the Copula likelihood takes its maximum are $(0.619,
0.241)$ at bin 1 and bin 2.

\begin{figure}[!t]
\begin{minipage}{.48\textwidth}
\begin{center}
\includegraphics[width=0.95\textwidth]{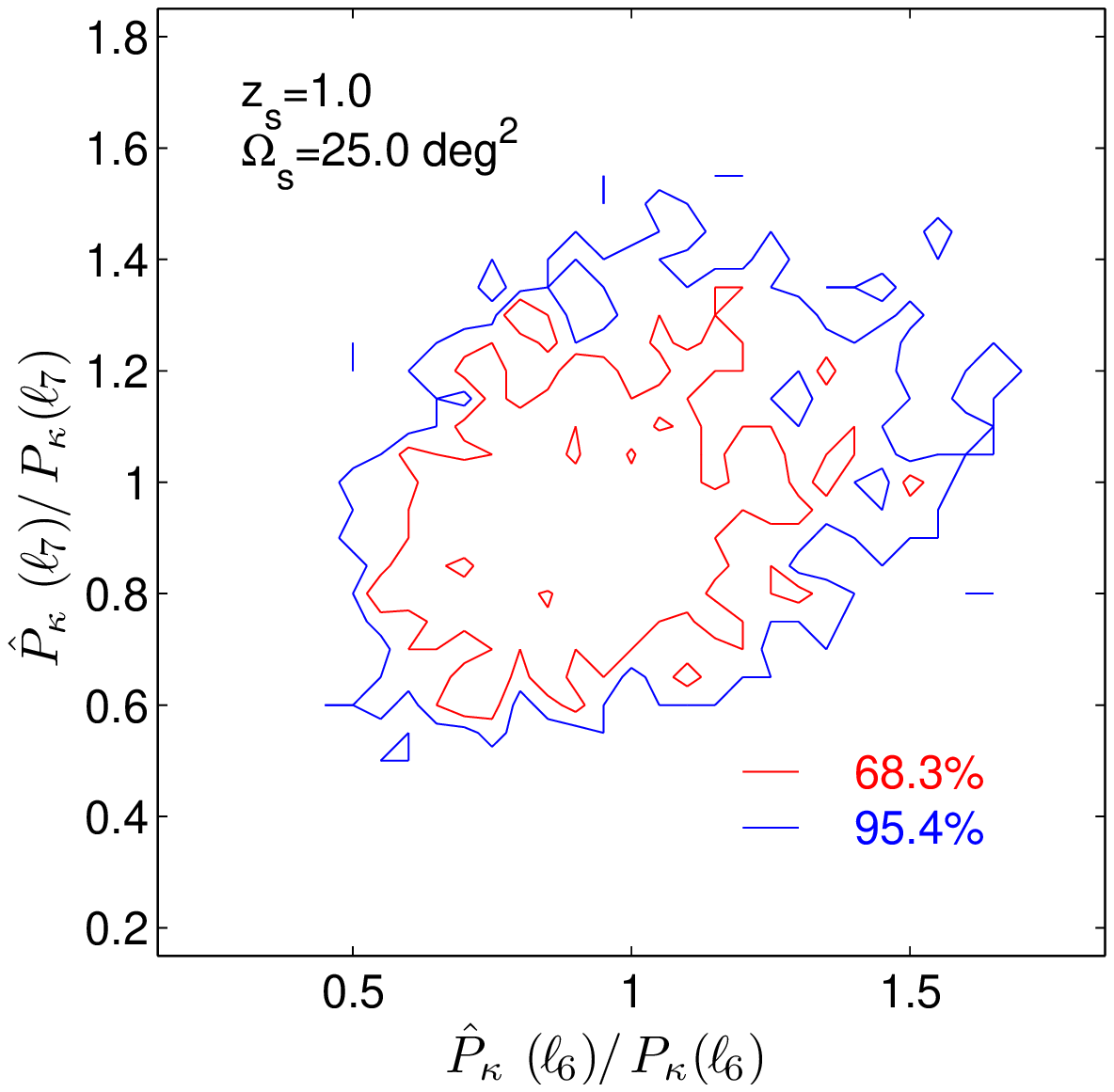}
\end{center}
\end{minipage}
\begin{minipage}{.48\textwidth}
\begin{center}
\includegraphics[width=0.95\textwidth]{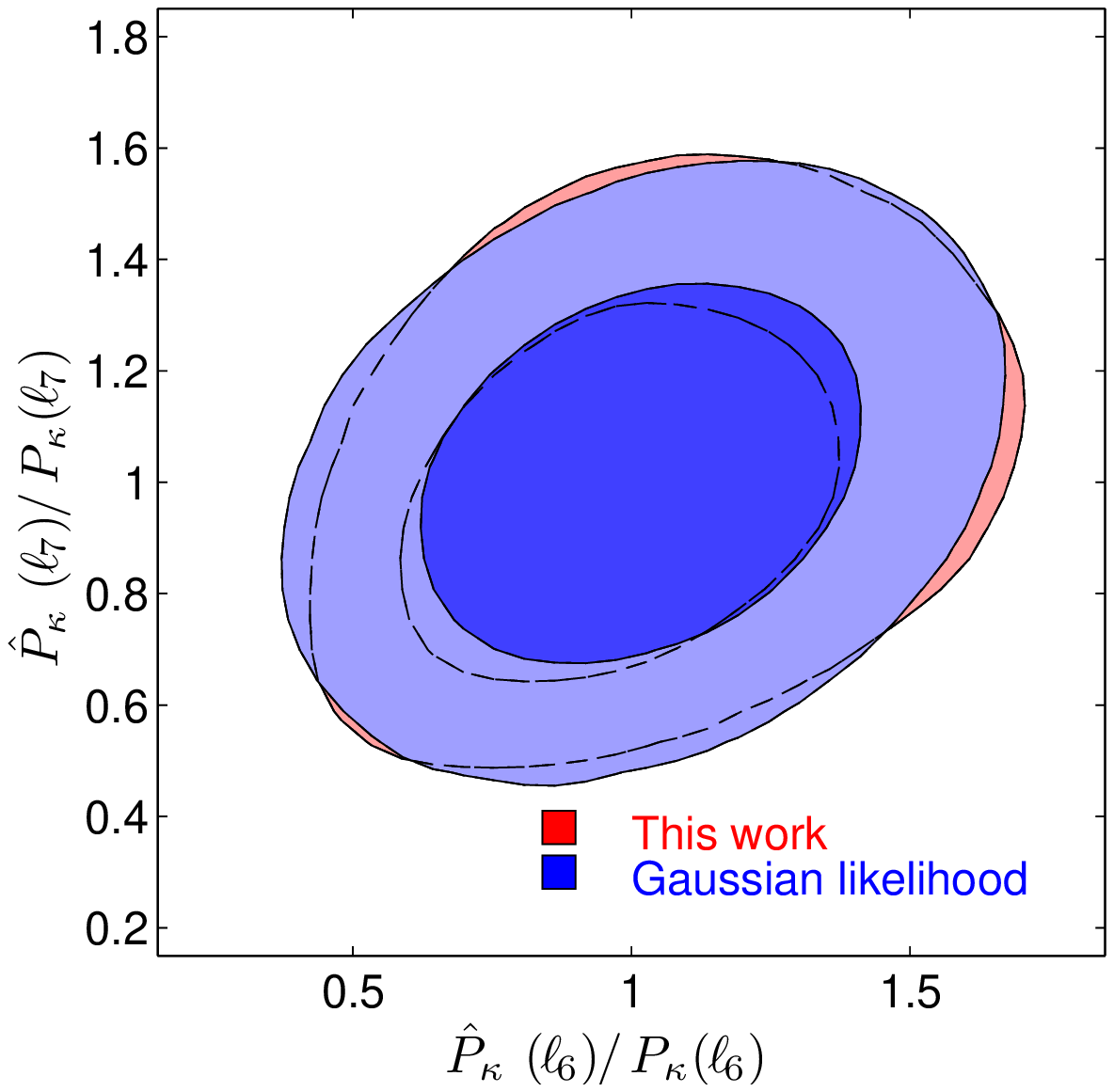}
\end{center}
\end{minipage}
\vskip-\lastskip
\caption{Same as Fig.~\ref{fig:2dim0102}, but convergence power spectrum
 is estimated at bin 6 and bin 7.}
\label{fig:2dim0607}
\end{figure}

Figure~\ref{fig:2dim0607} is same as Fig.~\ref{fig:2dim0102}, but 
results from bin 6 power and bin 7 power.
The bin 6 and bin 7 correspond to the multipoles $\ell=323$ and
$\ell=436$, respectively.
The results from our likelihood are similar to the results from the Gaussian
likelihood, because the $\chi^2$ distribution approaches Gaussian
distribution at these scales.
However, we can see the small deviation from each mean value between the
results obtained from the two likelihoods. 
 
\begin{figure}[!t]
\begin{minipage}{.48\textwidth}
\begin{center}
 \includegraphics[width=0.95\textwidth]{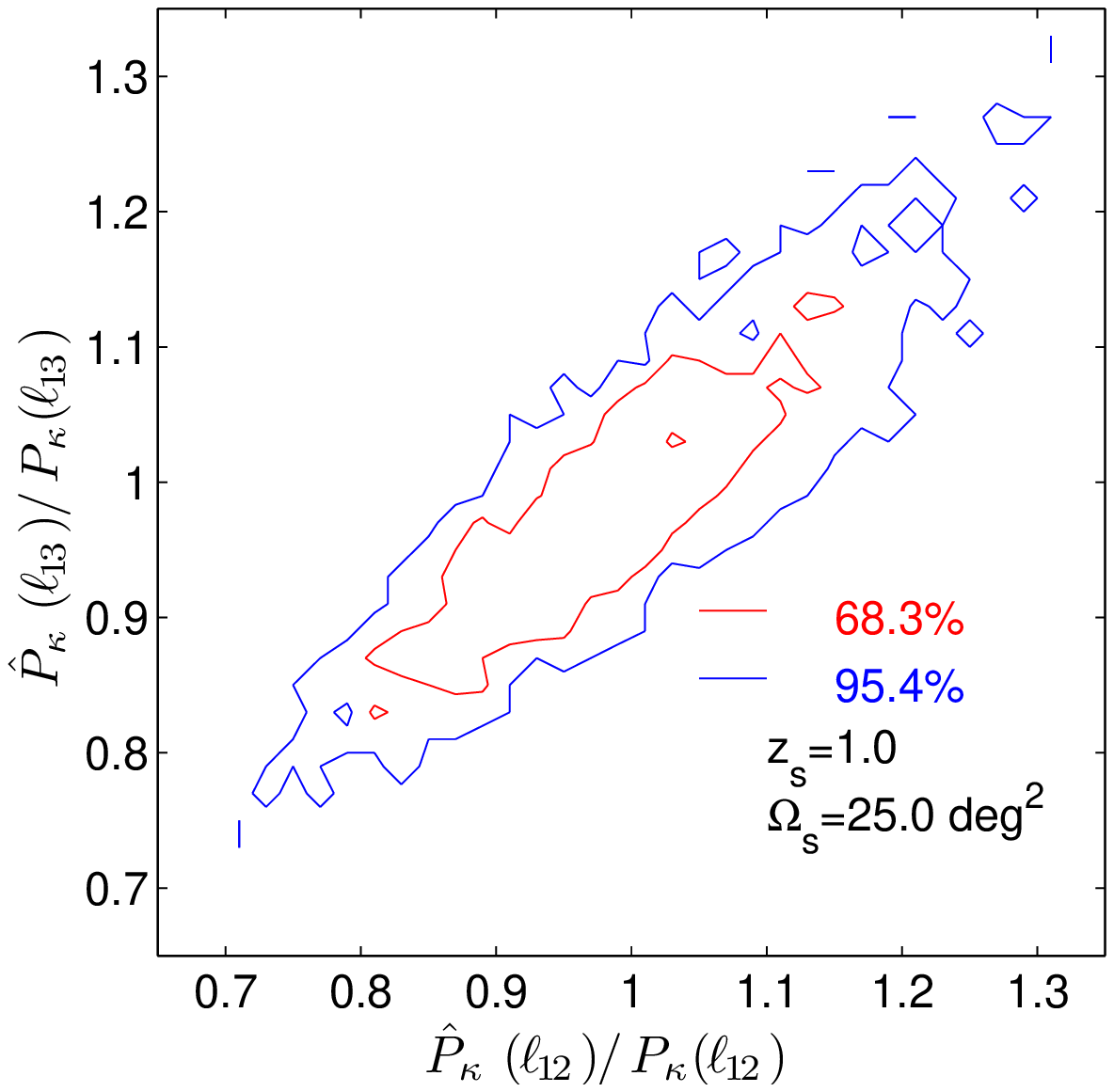}
\end{center}
\end{minipage}
\begin{minipage}{.48\textwidth}
\begin{center}
\includegraphics[width=0.95\textwidth]{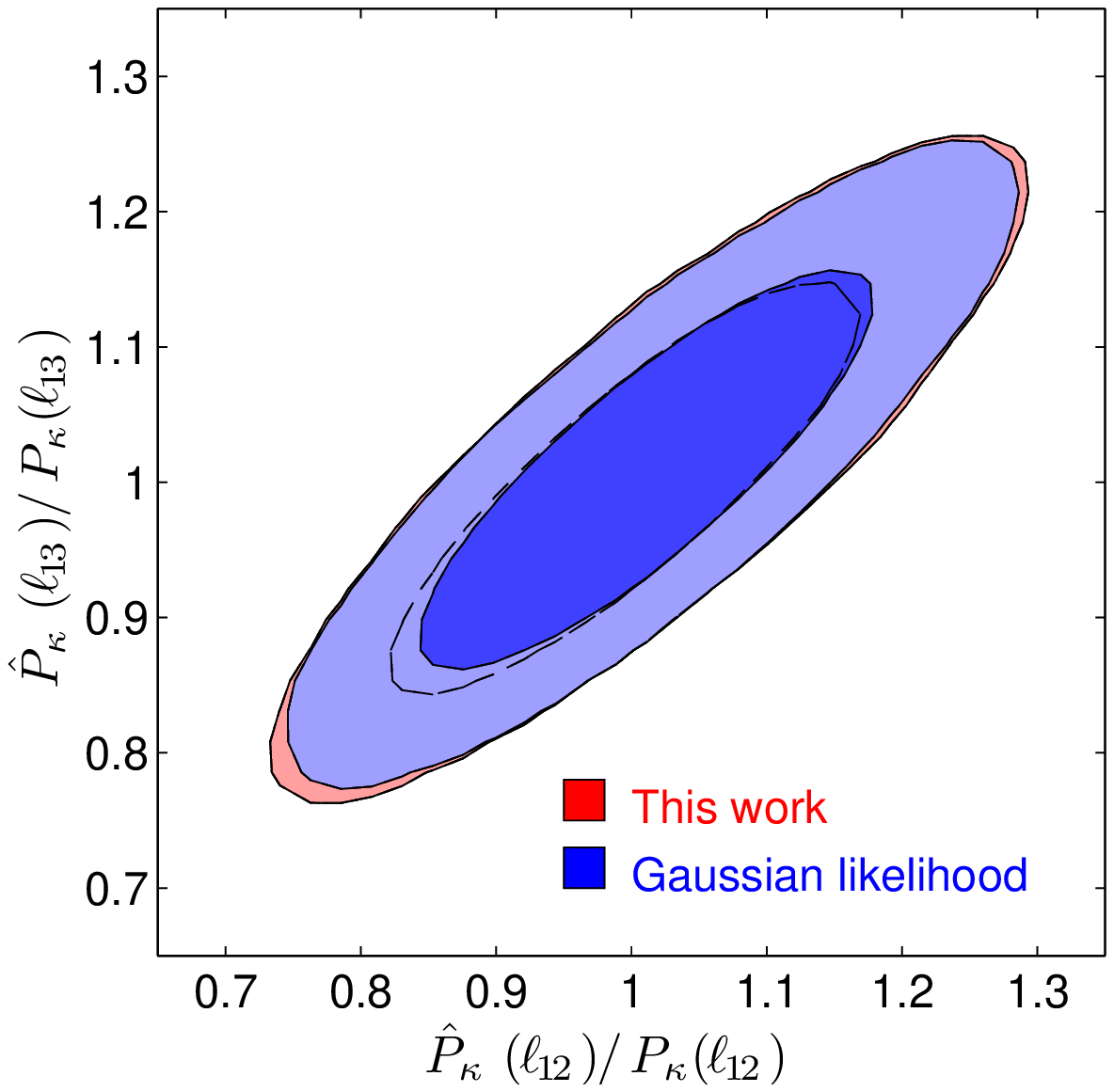}
\end{center}
\end{minipage}
\vskip-\lastskip
\caption{Same as Fig.~\ref{fig:2dim0102}, but convergence power spectrum
 is estimated at bin 12 and bin 13.}
\label{fig:2dim1213}
\end{figure}

Figure~\ref{fig:2dim1213} is also the same as Fig.~\ref{fig:2dim0102}, but results 
from bin 12 power and bin 13 power.
The bin 12 and bin 13 correspond to the multipoles $\ell=1952$ and
$\ell=2635$, respectively.
The contours from the copula likelihood are nearly identical to those from
the Gaussian likelihood, because the $\chi^2$ distribution approaches to
Gaussian
distribution at these small scales due to the central limit theorem.
We can see a positive strong correlation which is mainly attributed
to the fact that the nonlinear gravitational evolution causes the
non-Gaussian contribution and correlations between different multipole
bins (e.g., see in Fig. 7 in \citep{2009ApJ...701..945S}).

\begin{figure}[!t]
 \includegraphics[width=0.45\textwidth]{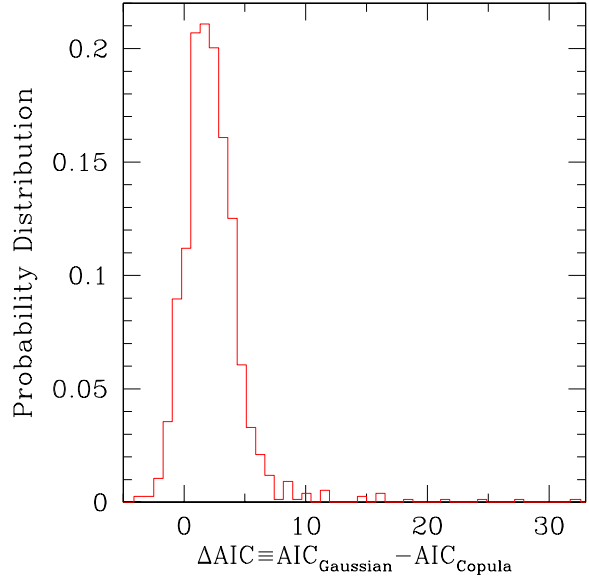}
\vskip-\lastskip
\caption{Probability distribution of the Akaike
 Information Criterion (AIC) difference between the Copula likelihood 
(Eq.~\ref{gau_copula_like}) and the Gaussian likelihood
 (Eq.~\ref{gau_like}), among 1000 realizations.
 }
\label{fig:pro_aic}
\end{figure}

To confirm a much better reliability of our Copula likelihood, we
examine how the Copula likelihood is {\em statistically} better relative
to the Gaussian likelihood based on the information criterion theory.
We use the Akaike Information Criterion (AIC) \citep{1974ITAC...19..716A,2000Ap&SS.271..213T,spiegelhalter2002bayesian}
to evaluate which model is more preferable.
The AIC is defined as 
\begin{equation}
{\rm AIC}\equiv -2\ln \mathcal{L}_{\rm max}+2k,
\end{equation}
where $\mathcal{L}_{\rm max}$ is the maximum likelihood achievable by
the model under a certain dataset, and $k$ is the number of free
parameters of the model.
In the present analysis $\mathcal{L}_{\rm max}$ is directly obtained by
substituting  $\hat{P}_\kappa(\ell_i)$ into Eqs.~(\ref{gau_copula_like})
and (\ref{gau_like}) at each realization, because
model parameters have been already fixed to give the maximum probability
for the 1000 realized simulation data.
The meaning of the AIC is clearly understood as an extension of the
maximum likelihood method.
An explanation for astronomers can be found in
\citep{2000Ap&SS.271..213T} and there are also useful applications of
the AIC in their paper.
Using the AIC enables us to compare the goodness of a certain model with
that of another type directly.
Figure~\ref{fig:pro_aic} shows the PDF of the AIC difference, 
 $\Delta{\rm AIC}\equiv{\rm AIC}_{\rm Gaussian}-{\rm AIC}_{\rm Copula}$
 among 1000 realizations. 
We see that the AIC is significantly
reduced by the Copula likelihood (positive $\Delta {\rm AIC}$) for 
the vast majority of the realizations.
This means that the Copula likelihood
function represents the distribution of the convergence power spectra
from Monte Carlo simulations better than the Gaussian likelihood one.
We estimate that the mean value of the AIC difference is 2.35. 
Some rough rules of thumb are available and useful for estimation of 
the goodness of models. As shown in
\citep{burnham2002model}, the mean value of the AIC difference we have
estimated is regarded as {\em Considerably less} which means the evidence 
to support the Gaussian likelihood is considerably less than the
Copula likelihood. 
Therefore we can conclude that our Copula likelihood is strongly favored
compared to the Gaussian likelihood for cosmic shear power spectrum.

Finally, we discuss how the difference between the two likelihoods
affects the cosmological parameter estimation, paying attention to the
number of bins.  
Since weak lensing power spectra are expected to be
very smooth in $\ell$, it is very likely that little information is lost by binning as
long as the bins are narrow compared to the width of any features.  If
we consider an observation over the sky coverage $f_{\rm sky}$
($f_{\rm sky}=\Omega_{\rm s}/4\pi$) with useful signal at
$\ell_{f}\simlt\ell\simlt\ell_{\rm max}$, the total number of modes one
can obtain is estimated to be $n_{t}\sim f_{\rm sky}(\ell_{\rm
max}^2-\ell_{f}^2)$.  If the number of bins is small enough, the
distribution of the power at each bin can be approximated by a Gaussian
because a sufficient number of modes are in that bin.  As is shown
in \citep{2008PhRvD..77j3013H} if the number of bins satisfies a
condition $n_{b}\ll f_{\rm sky}^{1/2}\ell_{\rm max}$ assuming $\ell_{\rm
max}^2\gg\ell_{f}^2$, the Gaussian approximation for the one-point PDF at
each bin becomes very good.  In that case one can use the
multivariate Gaussian likelihood in order to estimate the cosmological
parameters, which will simplify the parameter estimation analysis.  Now
let us apply this argument to our examples. In our example with
$f_{\rm sky}=6.25\times 10^{-4}$ and $\ell_{\rm max}\sim 1000$, the
square root of the 
total number of the modes is $n_t \sim f_{\rm sky}^{1/2}\ell_{\rm
max}=25$.  This number is not so large compared to the number of bins,
i.e., $n_b=13$.  Therefore, in this case an accurate likelihood function
should be used in order to get unbiased cosmological parameter
constraints, instead of the conventional multivariate Gaussian
likelihood \citep[see,][]{2010PhRvL.105y1301S}.  
Meanwhile, if we consider a future type survey which has $f_{\rm sky}=0.05$ and
$\ell_{\rm max}\sim 1000$ with the same number of bins, the square root
of the total number
of the modes is $f_{\rm sky}^{1/2}\ell_{\rm max}=224$ which is much
larger than the number of bins.  Then, the Gaussian approximation could
be fine in this case  \citep[][]{2010PhRvL.105y1301S} \footnote{This is a rough
discussion because our binning is logarithmic in $\ell$ space and
each bin does not contain an equal number of the modes.}.   Note, however,
that Gaussian approximation may be violated by natural binning which has
optimal $\ell$ resolution, regardless of how $\ell_{\rm max}$ value is.
Therefore, we suggest the Copula likelihood should be used when an optimal
binning is done to keep the cosmological information as much as
possible.  The impact of the difference between the two
likelihoods on cosmological parameter estimations is illustrated and discussed
carefully in a companion paper \citep{2010PhRvL.105y1301S}.

\section{Conclusion}
It has been becoming an important issue to obtain not only accurate
statistics such as a power spectrum and covariance matrix but also an
accurate likelihood function for the precision cosmology.  In this work, we
introduced a statistical tool called a copula into cosmology in a rather pedagogical
way. The copula is a function to generate an $n$-point CDF from the given
one-point CDFs and prescribed dependence structure of variables. 
We then applied the copula to the cosmic convergence power
spectrum estimated from 1000 realizations obtained from ray-tracing
simulations generated by \citep{2009ApJ...701..945S}.

By taking into account the fact that the one-point PDF of the
convergence power spectrum is well approximated by $\chi^2$
distribution, we showed that the Gaussian copula can reproduce the
$n$-dimensional shape of the likelihood function of the convergence
power spectrum better than the multivariate Gaussian likelihood, for the
simulated 13-binned data expected from the assumed survey area of
$\Omega_{\rm s}=25$ deg$^2$.  This is a main result of the present
paper. The differences between the two likelihood models become
significant at lowest multipole bins.

The deviation from the multivariate Gaussian will heavily depend on the
width of the binning. We discussed that a sufficiently sparse
binning will make it possible to use a multivariate Gaussian likelihood
for the future cosmic shear survey. However, if one makes an optimal
binning ($\Delta \ell \sim 1$) to keep the information as much as
possible, non-Gaussian corrections will become important. In this case
the copula will provide us an appropriate likelihood function in a
convenient way by relating the one-point CDFs to the $n$-point CDF.

\begin{acknowledgments}
We would like to thank the anonymous referees for careful reading of our
manuscript and very useful suggestions.
M.S is supported by the JSPS. 
T.T.T. has been supported by Program for
Improvement of Research Environment for Young Researchers from Special
Coordination Funds for Promoting Science and Technology.  This work is
partially supported by the Grant-in-Aid for the Scientific Research Fund
No. 20740105 (T.T.T.), No. 21740177, No. 22012004 (K.I.) and Grant-in-Aid for
Scientific Research on Priority Areas No. 467 ``Probing the Dark Energy
through an Extremely Wide and Deep Survey with Subaru Telescope''
commissioned by the MEXT of Japan.
\end{acknowledgments}

\bibliography{ms}

\clearpage

\end{document}